\documentclass[floatfix,twocolumn,showpacs,amsmath,amssymb]{revtex4-1}
\usepackage{graphicx}
\usepackage{epsfig}
\usepackage{subfigure}
\usepackage{dcolumn}
\usepackage{bm}
\setcitestyle{super}

\begin{document}

\title{Effective mass versus band gap in graphene nanoribbons: influence of H-passivation and uniaxial strain}
\author{Benjamin O. Tayo}
 \affiliation{Physics Department, Pittsburg State University, Pittsburg, KS 66762, USA}

\begin{abstract}
A simple model which combines tight-binding (TB) approximation with parameters derived from first principle calculations is developed for studying the influence of edge passivation and uniaxial strain on electron effective mass of armchair graphene nanoribbons (AGNRs). We show that these effects can be described within the same model Hamiltonian by simply modifying the model parameters i.e., the hopping integrals and onsite energies. Our model reveals a linear dependence of effective mass on band gap for H-passivated AGNRs for small band gaps. For large band gap, the effective mass dependence on band gap is parabolic and analytic fits were derived for AGNRs belonging to different families. Both band gap and effective mass exhibit a nearly periodic zigzag variation under strain, indicating that the effective mass remains proportional to the band gap even when strain is applied. Our calculations could be used for studying carrier mobility in intrinsic AGNRs semiconductors where carrier scattering by phonons is the dominant scattering mechanism.
\end{abstract}
\maketitle
\section{Introduction}
Graphene is a two-dimensional (2D) allotrope of carbon with excellent electronic and mechanical properties, making it suitable for multiple applications in nanoscale electronics and nanophotonics.\cite{Novoselovone, Novoselovtwo} A major deficiency in graphene's properties is the absence of a band gap rendering it impossible for use in switching circuits.\cite{Geim} Several approaches have been used to induce a band gap in graphene such as electrically gated bilayer graphene,\cite{Castro,Zhang,McCann} substrate induced band gap,\cite{Zhou, Gionannetti} or isoelectronic codoping with boron and nitrogen.\cite{Lei} Recently, it has become possible to engineer the band gap of graphene by lithographic patterning into small quasi one-dimensional (1D) nano sheets referred to as graphene nanoribbons (GNRs)\cite{Hanone, Hantwo, Todd} with excellent electronic properties such as room temperature ballistic transport.\cite{Baringhaus,Palacios} The ability to produce GNRs in very large amount is helping to accelerate research in the field of GNR electronics. As quasi 1D materials, GNRs are extremely sensitive to their surrounding conditions, which provides a route for manipulating their electronic properties. Additionally, other factors such as finite size effect,\cite{Hod,Nakada} edge effect,\cite{Yson,Lee,Sodi,Gorjizadeh,Simbeck,Wang} and the presence of strain\cite{Xihong,Li,Lu} could be used to effectively tune the electronic properties GNRs.

While patterning graphene into GNRs helps to induce a band gap, it has been shown that the effective mass of GNRs is proportional to the band gap\cite{jinyang wang}. In intrinsic semiconductors where carrier scattering by longitudinal acoustic phonons is the dominant scattering mechanism, the mobility is inversely proportional to the effective mass\cite{bardeen, belenaz}. This means increasing the band gap of GNRs results in an increase in effective mass and a decrease in mobility\cite{ouyang}. It is therefore extremely important to perform a comprehensive study on the influence of edge passivation and external strain on carrier effective mass and its correlation with the energy band gap.

The combined effects of edge passivation and strain on band gap for GNRs has been extensively studied using first principle calculations\cite{Xihong}. In Ref. 23, the authors focused only on modulation of band gap due to edge effects and strain. In Ref. 26, it is shown that the effective mass of AGNRs is proportional to the band gap, but the authors did not take into account the bond length variation induced by H-passivation.

In this work, we present a simple model which combines TB approximation with parameters derived from first principle calculations for studying the influence of edge passivation and uniaxial strain in the -16\% to 16\% range on the effective mass of AGNRs. We show that these effects can be described within the same model by simply changing the model parameters like the hopping integrals and onsite energies. For unstrained H-passivated AGNRs, we show that the effective mass displays a linear dependence on energy band gap for small band gap energies. For large band gap, the effective mass dependence on band gap is parabolic and analytic fits were derived for AGNRs belonging to different families. Both band gap and effective mass exhibit a nearly periodic zigzag variation under strain, indicating that the effective mass remains proportional to the band gap even when strain is applied. Our analysis explains in a simple and computationally very efficient way, the physical mechanism that gives rise to the significant modulation of the electronic properties of GNRs and is useful for modelling charge transport in graphene nanoribbons.

This paper is organized as follows: In Sec. \ref{two}, we describe the general formalism. In Sec. \ref{results}, we discuss the results. In Sec. \ref{dos},
we study the density of quantum states in the presence of strain. A short summary concludes the paper.
\section{General Formalism}\label{two}
\begin{figure}[t]
\centering
\includegraphics[width=3.6 in]{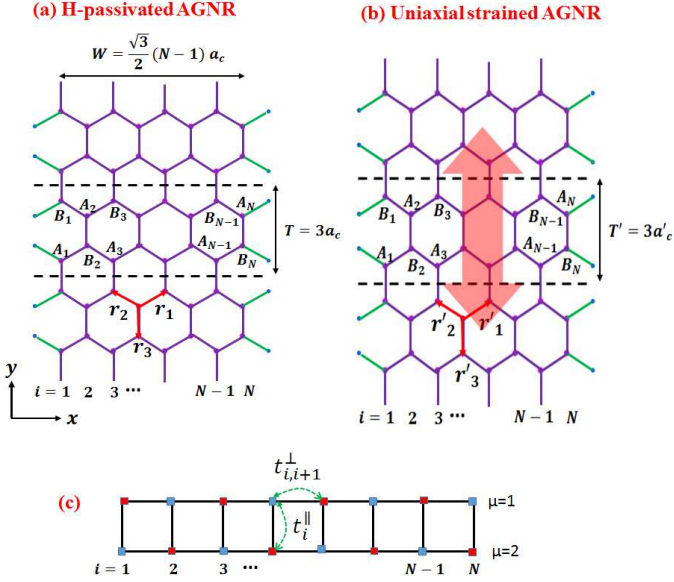}
\caption{(a) Unstrained H-passivated AGNR showing the number of dimer lines along the width of the ribbon. (b) H-passivated AGNR under uniaxial strain. (c) Two-leg ladder with $N$ rings representing the equivalent TB Hamiltonian of the system at the $\Gamma$ point. Within our model, systems (a) and (b) are described by the same Hamiltonian matrix (c) by simply modifying the hopping integrals $t^{\parallel}_{i} $ and $t^{\perp}_{i,i+1}$.}
\label{AGNR structure}
\end{figure}

Carrier scattering  by longitudinal acoustic phonons plays a significant role in charge transport in intrinsic semiconductors. Within the deformation potential theory, the relaxation time ($\tau_{dp}$) and mobility $(\mu_{dp})$ for electron-phonon ($\tau_{dp}$) scattering for quasi-1D systems like AGNRs is given as\cite{belenaz}
\begin{eqnarray}\label{relaxation time}
\tau_{dp} = {\hbar^2 C \over (2\pi k_B T)^{1/2}(m^{*})^{1/2}E_1^{2}}, ~~\mu_{dp} = {e\tau_{dp} \over m^{*}}
\end{eqnarray}
where the subscript ``dp" stands for deformation potential, $T$ is the temperature, $C$ is the longitudinal elastic constant, $m^{*}$ the effective mass, and $E_1$ the deformation potential constant reflecting the change in band edge induced by strain. While patterning graphene into nanoribbons and the application of external strain could be employed to induce and manipulate the band gap of graphene, the effective mass of GNRs has been shown to vary proportionately with the band gap for perfectly terminating AGNRs\cite{jinyang wang}, accounting for the diminishing mobility with increasing band gap in graphene nanoribbon field effect transistors\cite{ouyang}. It is therefore extremely important to perform a comprehensive study on the influence of edge passivation and external strain on carrier effective mass and its correlation with the energy band gap in GNRs.

We consider an AGNR of width $W = {\sqrt{3}\over 2 }(N-1)a_c$ and translation period  $T = 3a_c$, where $N$ is the  number of dimer lines and $a_c \sim 1.423~\mathrm{\AA}$ the unstrained carbon to carbon (C-C) bond length at the center of the GNR (see Fig. \ref{AGNR structure}(a)). Since the width of an AGNR is specified by the number of dimer lines along the ribbon, we will use the notation N-AGNR to refer to an AGNR with N dimer lines along the ribbon. The unit cell of an N-AGNR contains $N$ A-type atoms and $N$ B-type atoms, as shown in Fig. \ref{AGNR structure}. Additionally,  N-AGNRs can be classified into three distinct families $N = 3p, 3p+1 , 3p+2$, where $p$ is a positive integer and their electronic properties are known to exhibit distinct family splitting.\cite{Wakabayashi,Ezawa,Brey,Sasaki,Abanin} The dangling $\sigma$-bonds at the edges are passivated by H atoms (or other atoms/groups like O and OH). Edge passivation by foreign atoms or groups produces geometric deformation altering the C-C bonds and bonding angles at the nanoribbon edge. \cite{Wang,Hosoya,Fujita} For example, for AGNRs passivated with H atoms, the bond lengths parallel to dimer lines at edges are shortened by about 3.5\%,\cite{Yson} compared to those in the middle of the ribbon. In general, this kind of geometric deformation results in changes of the hopping parameter\cite{Porezag} between two neighboring carbon atoms and onsite energies on the GNR edge. In Fig. \ref{AGNR structure} (b), we show the H-passivated AGNR under uniaxial strain. In the presence of uniaxial strain, the translational period becomes $T' = 3a'_{c}$, where $a'_c$ is the bond length for AGNR under strain. Hence, the strain ($\sigma$) can be defined as $\sigma = (a'_c - a_c)/a_c$. A positive value for $\sigma$ corresponds to tensile strain while a negative value represents compressive strain. Since edge passivation and the presence of strain both alter the C-C bond length, these two effects can be described within the same model by simply incorporating the changes in onsite energies and hopping integrals induced by these effects. We shall discuss these effects using the TB model in what follows.

The electronic states of GNRs are expressed in terms of the axial momentum ($k$) and the lateral momentum ($k_n$), where $n$ in an integer describing the quantization of the component of electron's momentum along the width of the ribbon. AGNRs are semiconductors with a direct band gap at the $\Gamma$ point. At $k=0$, the TB  Hamiltonian for an AGNR reduces to a two-leg ladder lattice system \cite{Yson}, as shown in Fig. \ref{AGNR structure} (c). The Hamiltonian of this simpler model reduces to
\begin{eqnarray}\label{lattice model hamiltonian}
\mathcal{H} &=& \sum_{i=1}^{N}\sum_{\mu =1}^{2} \varepsilon_{\mu,i}a^{\dagger}_{\mu,i}a_{\mu,i}  -\sum_{i=1}^{N-1} \sum_{\mu =1}^{2} t^{\bot}_{i,i+1}(a^{\dagger}_{\mu,i+1}a_{\mu,i} + \nonumber \\
&& h.c.)-\sum_{i=1}^{N} t^{\|}_{i}(a^{\dagger}_{1,i}a_{2,i} + h.c.)
\end{eqnarray}
where $({i,\mu})$ denote a site, $\varepsilon_{\mu,i}$ site energies, $t^{\bot}_{i,i+1}$ and  $t^{\|}_{i}$ the
nearest neighbor hopping integrals within each leg and between the legs respectively, and $a_{\mu,i}$ the annihilation
operator of $\pi$-electrons on the $i$-th site of the $\mu$-th leg. We remark here that in this model, the electronic properties of GNRs are sensitive only to the three parameters: the site energies $\varepsilon_{\mu,i}$, and the nearest neighbor hopping integrals $t^{\bot}_{i,i+1}$ and $t^{\|}_{i}$. These parameters will differ for perfectly terminating, edge passivated
and strained GNRs. This means that the combined effects of edge passivation and strain can be described by the same model Hamiltonian by modifying the TB parameters in order to account for the considerable changes in C-C bond lengths. Thus, the model is very simple and computational very efficient.
In general for $k \neq 0$, $\mathcal{H}$ can be expressed in matrix form for the translationally invariant system. If we order the basis as $A_1$, $B_2$, $A_3$, $\dots$, $A_{N-1}$, $B_{N}$, and $B_1$, $A_2$ , $B_3$, $\dots$, $B_{N-1}$, $A_N$, then the nearest neighbor Hamiltonian can be split into four $N\times N$ blocks
\begin{eqnarray}\label{lattice hamiltonian block}
\mathcal{H}(k) = \left(\begin{array}{cc}
                      \mathcal{H}_1 & \mathcal{H}_{12} \\
                     \mathcal{H}_{12}^{\dagger} & \mathcal{H}_{2}
                   \end{array}\right)
\end{eqnarray}
where
\begin{eqnarray}\label{lattice block one}
\mathcal{H}_1 &=& \left(\begin{array}{cccc}
                      \varepsilon_{1,1} & t^{\bot}_{1,2}&0 & \dots \\
                      t^{\bot *}_{1,2}& \varepsilon_{1,2} & t^{\bot}_{2,3}&\dots  \\
                      0&  t^{\bot *}_{2,3} &\varepsilon_{1,3}& \dots \\
                      \dots & \dots & \dots & \dots
                   \end{array}\right) \nonumber\\
 \mathcal{H}_2 &=& \left(\begin{array}{cccc}
                      \varepsilon_{2,1} & t^{\bot}_{1,2}&0 & \dots \\
                      t^{\bot *}_{1,2}& \varepsilon_{2,2} & t^{\bot}_{2,3}&\dots  \\
                      0&  t^{\bot *}_{2,3} &\varepsilon_{2,3}& \dots \\
                      \dots & \dots & \dots & \dots
                   \end{array}\right)\nonumber \\
\mathcal{H}_{12} &=& \left(\begin{array}{cccc}
                      t^{\|}_{1} d_k & 0&0 & \dots \\
                     0&  t^{\|}_{2}  & 0&\dots  \\
                      0&  0 & t^{\|}_{3} d_k& \dots \\
                      \dots & \dots & \dots & \dots
                   \end{array}\right)\nonumber \\
\end{eqnarray}
Here, $d_k = e^{-i kT}$, with $T$ being the lattice constant. The electronic band structure of the AGNR can then be obtained by solving the eigenvalue equation
\begin{eqnarray}\label{lattice model hamiltonian}
\mathcal{H}(k) \mathbf{\mathcal{C}}_{\lambda n}(k) =  E_{\lambda n}(k)\mathbf{\mathcal{C}}_{\lambda n}(k)
\end{eqnarray}
where $\lambda = c ~(v) $ corresponds to the conduction (valence) band, and $n$ the band index. The coefficients $\mathbf{\mathcal{C}}_{\lambda n}(k)$ are TB wave function amplitudes.

Edge passivation and the effect of strain can both be described within our model by modifying the onsite energies and hopping integrals. Strains applied to the GNR and the absorption of atoms or molecules at the edges causes an increase or decrease in C-C bond lengths, which in turn alters the onsite energies and the hopping integrals. Previous studies carried out for edge passivated GNRs have shown that to first-order, the change in onsite energy due to edge passivation does not alter the band gap.\cite{Yson,Wang} We will therefore assume that changes in onsite energies due to H-passivation and uniaxial strain are negligible. Hence we shall set all the onsite energies at $ \varepsilon_{\mu,i} = 0$ for $\mu = 1,2$ and $n = 1,2, \dots, N$. In our treatment, we then focus only on changes in the hopping integrals due to external perturbations.

A decrease in C-C bond length will increase overlap of $\pi$ orbitals which leads to an increase in the hopping integral. Likewise, an increase in bond length will result to a decrease in $\pi$ orbital overlap, which accordingly decreases the hopping integral. The analytic expressions for TB matrix elements between carbon atoms as a function of the C-C bond length can be expressed in terms of the Chebyshev polynomials $T_m(x)$  yielding\cite{Porezag}
\begin{eqnarray}\label{carbon to carbon hopping integral}
H_{\pi}^{CC}(r) = \sum_{m=1}^{10} c_{m} T_{m-1}(y) -{c_1 \over 2}, ~~ y = {r - {b+a \over 2} \over {b-a \over 2}}
\end{eqnarray}
where $r \in (a, b)$ is the interatomic distance for C-C interactions, and  $(a, b)$ the range of values over which the expansion is valid. The coefficients $c_m$ and boundaries $a$ and $b$ are tabulated in Ref. 37.
\begin{table}[h]
\caption{TB parameters for AGNRs under uniaxial strain. Parameters were calculated using an unstrained C-C distance of $a_c =1.423~\mathrm{\AA}$ and $t=2.7$ eV.}
\label{TB parameters for strained GNRS}
\begin{tabular}{ccccc}
\hline
\hline
$\sigma$ & $a^{\|}_{c}$ (\AA) & $t^{\|} (eV) $&  $a^{\bot}_{c}$ (\AA) & $t^{\bot} (eV) $ \\
\hline
\hline
-0.16&	1.195&	4.506&	1.340&	3.256\\
-0.15&	1.210&	4.365&	1.345&	3.219\\
-0.14& 1.224&	4.229&	1.350&	3.183\\
-0.13&	1.238&	4.097&	1.356&	3.147\\
-0.12&	1.252&	3.969&	1.361&	3.111\\
-0.11&	1.266&	3.844&	1.366&	3.075\\
-0.10&	1.281&	3.724&	1.371&	3.040\\
-0.09&	1.295&	3.607&	1.376&	3.005\\
-0.08&	1.309&	3.493&	1.381&	2.970\\
-0.07&	1.323&	3.383&	1.386&	2.935\\
-0.06&	1.338&	3.276&	1.391&	2.901\\
-0.05&	1.352&	3.173&	1.397&	2.867\\
-0.04&	1.366&	3.072&	1.402&	2.833\\
-0.03&	1.380&	2.975&	1.407&	2.799\\
-0.02&	1.395&	2.880&	1.412&	2.766\\
-0.01&	1.409&	2.789&	1.418&	2.733\\
0.00&	1.423&	2.700&	1.423&	2.700\\
0.01&	1.437&	2.614&	1.428&	2.667\\
0.02&	1.451&	2.530&	1.434&	2.635\\
0.03&	1.466&	2.449&	1.439&	2.603\\
0.04&	1.480&	2.371&	1.444&	2.571\\
0.05&	1.494&	2.294&	1.450&	2.540\\
0.06&	1.508&	2.221&	1.455&	2.509\\
0.07&	1.523&	2.149&	1.461&	2.478\\
0.08&	1.537&	2.079&	1.466&	2.447\\
0.09&	1.551&	2.012&	1.472&	2.416\\
0.10&	1.565&	1.947&	1.477&	2.386\\
0.11&	1.580&	1.883&	1.483&	2.356\\
0.12&	1.594&	1.822&	1.488&	2.327\\
0.13&	1.608&	1.762&	1.494&	2.297\\
0.14&	1.622&	1.704&	1.499&	2.268\\
0.15&	1.636&	1.648&	1.505&	2.239\\
0.16&	1.651&	1.594&	1.510&	2.211\\
\end{tabular}
 \end{table}
 \section{Results and Discussion}\label{results}
We begin by calculating the band gap and electron effective mass for unstrained H-passivated AGNR. For perfectly terminating AGNR, we will set the nearest neighbor C-C TB hopping integral to $ t = 2.7$ eV, a value that has been used to successfully describe the electronic properties of graphene \cite{Reich}. For H-passivated AGNRs, the bond lengths parallel to dimer lines at edges are compressed by about $3.5 \%$ as compared to those in the middle of the ribbon. Using Eq. (\ref{carbon to carbon hopping integral}), we can show that a $3.5 \%$ compressive strain on the bond length at the edges induces a $12 \%$ increase in the hopping integral. The effect of H-passivation can then be accounted for by setting $t^{\|}_{i} = 3.024 $ eV, for $i=1$ and $i=N$, $t^{\|}_{i} = 2.7 $ eV for $i=2, \dots, N-1$, and $t^{\bot}_{i,i+1} = 2.7$ eV for $\mu =1,2$, $i = 1, \dots N-1$.  Substituting these parameters into Eq. (\ref{lattice hamiltonian block}) and diagonalizing the resulting Hamiltonian matrix,  we obtain the energy band structure of the AGNR. The electron effective mass $m_e$ for the lowest conduction band is obtained from the fit $ E_{c ,1}(k) = E_{c} + {\hbar^2 k^2 \over 2m^{*}}$, where $E_c$ is the conduction band edge. The hole effective mass is equal to electron effective mass for both unstrained and strained H-passivated AGNRs. In our approach, we neglect the change in band structure of the AGNR due to quasiparticle effects \cite{LiYang, Prezzi}.

\begin{figure}[t]
\centering
\includegraphics[width=3.2 in]{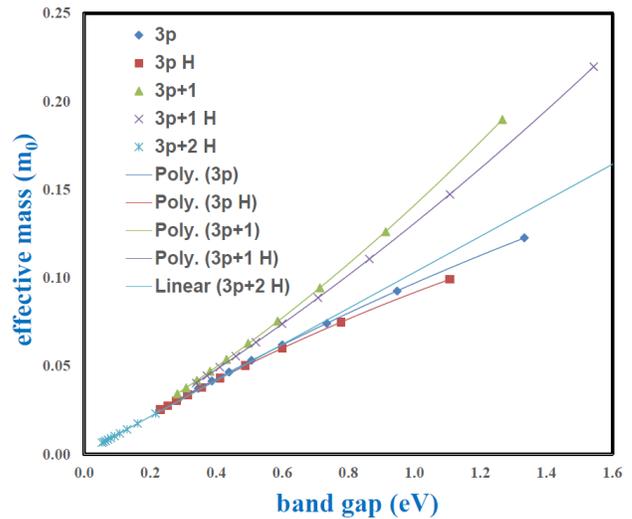}
\caption{Effective mass vs band gap for AGNRs with N = 6 - 35. The band gap is calculated at the $\Gamma$ point. The effective mass is computed from a parabolic fit of the lowest lying conduction band around the $\Gamma$ point. Effective mass shows distinct family splitting dependence on band gap, increasing with increasing band gap energy. The linear fit of the H-passivated $3p+2$ family represents the linear dependence of $m^{*}$ on $E_g$ at small band gaps. The other solid lines are parabolic fits for the $3p$ and $3p+1$ AGNRs. The effect of H-passivation is very prominent, increasing with increasing band gap or small ribbon width.}
\label{unstrained band gap and effe mass}
\end{figure}

In Fig. \ref{unstrained band gap and effe mass}, we show a plot of $m^{*}$ (in units of the free electron mass $m_0$) versus the energy band gap ($E_g$) for unstrained AGNR with $N = 6 - 35$. $m^{*}$ shows a distinct family splitting dependence on $E_g$, increasing with increasing $E_g$. $m^{*}$ varies between $0.006$ to 0.22 $m_0$ and obeys the same hierarchical pattern as $E_g$,\cite{Yson} with $m^{*} (3p+1)>m^{*} (3p)>m^{*}(3p+2)$ for all $p$. The effect of hydrogen passivation is also prominent, increasing with increasing band gap or decreasing ribbon width. Generally for a given band gap, the effective mass of the H-passivated AGNR is smaller than that of the unpassivated AGNR. In the TBA, the $3p+2$ AGNRs are gapless if the effect of H-passivation is not taken into account, that is why the figure only shows the effective mass of H-passivated $3p+2$ AGNRs. The variation of $m^{*}$ with $E_g$ can be fitted with the parabolic function
\begin{eqnarray}\label{parabolic}
m^{*}=  E_g (A + B E_g)
\end{eqnarray}
with $A$ and $B$ being the fitting parameters (see Tab. \ref{fitted parameters}). When $E_g$ is very small, $m^{*} \simeq  A E_g $. Using the analytic expression for the band energy\cite{jinyang wang}
\begin{eqnarray}\label{analytic band energy}
E(k)=  \pm \hbar v_F \sqrt{k^2 + \bigg({E_g\over 2\hbar v_F}\bigg)^2}
\end{eqnarray}
we find that
\begin{eqnarray}\label{mass fermi}
m^{*}=  {1 \over 2 v^{2}_F} E_g
\end{eqnarray}
or
\begin{eqnarray}\label{mass fermi}
A =  {1 \over 2 v^{2}_F}
\end{eqnarray}
from which we can estimate the Fermi velocity $v_F \approx 0.89 \times 10^6~\mathrm{m/s}$ .

\begin{table}[h]
\caption{Fitted parameters $A$ and $B$ for AGNRs belonging to distinct families.}
\label{fitted parameters}
\begin{tabular}{ccc}
\hline
family & $ A  (m_0/eV) $ & $ B  (m_0/eV^{2})  $ \\
\hline
\hline
3p &	0.1101&	-0.0142\\
3p H&	0.1098& -0.0192\\
3p+1 &  0.1053 & 0.0339\\
3p+1 H&	0.1071 & 0.0222\\
3p+2 H &	0.1022& 0.0000\\
\end{tabular}
 \end{table}

As shown in Fig. \ref{unstrained band gap and effe mass}, the $3p+2$ AGNRs have very narrow band gaps and small effective masses. However, in the presence of uniaxial strain, both the band gap and the electron effective mass gets significantly modulated, thus making it possible to engineer their electronic properties by applying strain.

We now consider the case of H-passivated AGNR under uniaxial strain. First we calculate the TB hopping integrals for non-passivated AGNR subjected to uniaxial strain, then we modify these parameters in order to take into account the effect of edge passivation. The unstrained bond vectors for an AGNR are given by (see Fig. \ref{AGNR structure} (a)):
\begin{eqnarray}\label{unstrained vectors}
\mathbf{r}_1 &=& a_c \bigg({\sqrt{3} \over 2} ~\hat{x} + {1 \over 2} ~\hat{y}\bigg)\nonumber \\
\mathbf{r}_2 &=& a_c \bigg(-{\sqrt{3} \over 2} ~\hat{x} + {1 \over 2} ~\hat{y}\bigg)\\
\mathbf{r}_3 &=& - a_c ~ \hat{y} \nonumber
\end{eqnarray}
where $\hat{y}$ is the axial direction of the AGNR. The application of a uniaxial strain
causes the following changes  (see Fig. \ref{AGNR structure} (b)):
\begin{eqnarray}\label{strained vectors}
\mathbf{r'}_1 &=& a_c \bigg[{\sqrt{3} \over 2}(1+ \nu\sigma) ~\hat{x} + {1 \over 2}(1+ \sigma) ~\hat{y}\bigg]\nonumber \\
\mathbf{r'}_2 &=& a_c \bigg[-{\sqrt{3} \over 2}(1+ \nu\sigma)~ \hat{x} + {1 \over 2}(1+ \sigma) ~\hat{y}\bigg]\\
\mathbf{r'}_3 &=& - a_c (1+ \sigma) ~ \hat{y} \nonumber
\end{eqnarray}
where $\sigma$ represents the uniaxial strain in the $\hat{y}$ direction, and $\nu \approx 0.165$ is the Poisson's ratio\cite{Blakslee, Yang}. Based on our model, which maps the AGNR to a two-leg ladder system, the C-C length in the axial direction ($a^{\|}_{c}$) and the corresponding length in the direction perpendicular to the axis ($a^{\bot}_{c}$) are given by
\begin{eqnarray}\label{strained C-C bond length}
a^{\bot}_{c}&=&|\mathbf{r'}_1|=|\mathbf{r'}_2| = a_c \bigg[\sqrt{3\bigg({1+ \nu\sigma \over 2}\bigg)^2 + \bigg({1+ \sigma\over 2}\bigg)^2}\bigg]\nonumber \\
a^{\|}_{c}&=&|\mathbf{r'}_3| =  a_c (1+ \sigma)
\end{eqnarray}
\begin{figure}[t]
\centering
\includegraphics[width=3.0 in]{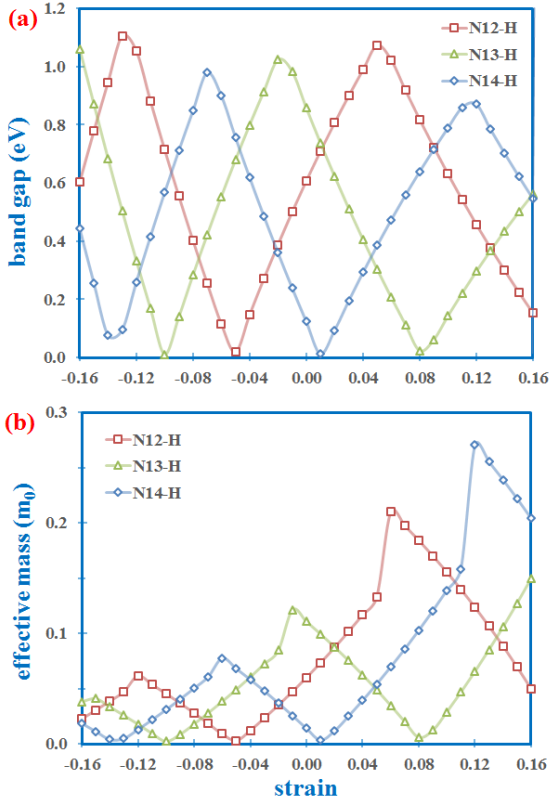}
\caption{(a) Bang gap and (b) effective mass as a function of strain for $N = 12, 13,$ and 14 H-passivated AGNRs. Notice that the hierarchical family pattern $m^{*}(3p+1)>m^{*} (3p)>m^{*} (3p+2)$ becomes invalid for strained H-passivated AGNRs. The zigzag pattern in both the band gap and effective mass indicates that the effective mass remains proportional to the band gap even in the presence of strain.}
\label{strained band gap effe mass}
\end{figure}
We can use the strained bond lengths  ($a^{\|}_{c}$) and ($a^{\bot}_{c}$) together with Eq. (\ref{carbon to carbon hopping integral}) to estimate the hopping integrals in the axial  ($t^{\|} $) and perpendicular ($t^{\bot} $) directions. These values are tabulated in Tab. \ref{TB parameters for strained GNRS} for a non-passivated AGNRs under uniaxial strain in the range -16\% to +16\%. As an example, for an AGNR under -16\% strain, we have  $t^{\|}_{i} = 4.506 $ eV and $t^{\bot}_{i,i+1} = 3.256$ eV. For H-passivated AGNRs under uniaxial strain, the bond lengths parallel to dimer lines at edges are compressed by an additional $3.5 \%$ compared to those in the middle of the ribbon (leading to an additional 12\% increase in hopping integral for the edge carbon atoms, as already discussed). This additional effect can be taken into account by setting $t^{\|}_{i} = 5.047 $ eV, for $i=1$ and $i=N$, $t^{\|}_{i} = 4.506 $ eV for $i=2, \dots, N-1$, and $t^{\bot}_{i,i+1} = 3.256$ eV for $\mu =1,2$, $i = 1, \dots N-1$. If we substituting these parameters into Eq. (\ref{lattice hamiltonian block}) and diagonalize the resulting Hamiltonian,  we obtain the energy band structure and effective mass for $\sigma = -16\%$. Applying the same process for strains in the -16\% to +16\% range allows us to successfully compute the band gap and effective mass for AGNRs under the combined effects of edge and strain.
We now apply our formalism to three AGNRs, namely N = 12, 13, and 14, representing the $3p, 3p+1$, and $3p+2$ families, respectively. In Fig. \ref{strained band gap effe mass}, we show the energy band gap and electron effective mass for H-passivated AGNRs under uniaxial strain in the range -16\% to 16\%.  Fig. \ref{strained band gap effe mass} (a) shows a zigzag pattern in the behavior of the band gap with strain for N = 12, 13, and 14 AGNRs. This pattern is due to changes in the C-C TB hopping integrals with C-C bond length when the AGNR is subjected to strain. \cite{Wang} The maximum value of the band gap for $N = 12$, 13, and 14 occur at +5\%, -2 \%, and -7\%, respectively, while the minimum value occur at -5\%, -10\%, and +1\%, respectively. These values agree extremely well with values obtained using first principle calculations and other results in the literature\cite{Xihong,Sun, Yang}. Notice also that the $3p+2$ AGNRs have very narrow band gaps, but when strain is applied, the band gap can be tuned up to about 1 eV. For instance, an N = 14 H-passivated AGNR has a band gap of only 0.123 eV in the absence of strain (see Fig. \ref{unstrained band gap and effe mass} (a)), but under a uniaxial strain of -7\%, the band gap becomes 0.979 eV (see Fig. \ref{strained band gap effe mass} (a)), which corresponds to about 700\% increase in the band gap.

Fig. \ref{strained band gap effe mass} (b) shows the electron effective mass plotted as a function of strain for the same AGNRs, which also exhibits a zigzag pattern but with peaks that increase as the applied strain changes from compressive to tensile. Uniaxial tensile strain thus have the tendency to increase the effective mass of an electron. The maximum value of the effective mass for $N = 12$, 13, and 14 occur at +6\%, -1 \%, and +12\%, respectively, while the minimum value occur at -5\%, -10\%, and +1\%, respectively. The minima of both $E_g$ and $m^{*}$ occur at the same values of strain while the maxima of $E_g$ and $m^{*}$ occur at different values of strain for the AGNRs considered. Similarly to the band gap, the electron effective becomes significantly modulated under the influence strain. For example, a H-passivated 12-AGNR has $m^{*} = 0.060~m_0$ in the absence of strain, but under a +6\% strain, the effective mass becomes $0.210~m_0$, which corresponds to a 250\% increase.

\section{Density of States}\label{dos}
\begin{figure}[t]
\centering
\includegraphics[width=3.1 in]{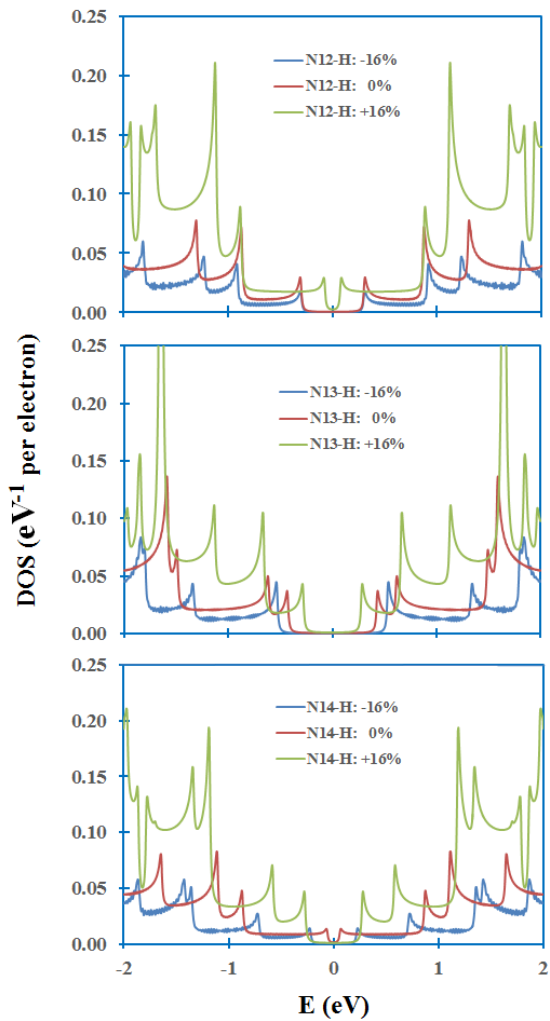}
\caption{DOS for $N = 12, 13,$ and 14 H-passivated AGNR for three values of strain: -16\%, 0\%, and +16\%.}
\label{densityofstates}
\end{figure}
It is very instructive to visualize the influence of uniaxial strain on electronic band structure by plotting the density of states (DOS). The finite
temperature DOS per electron is given by \cite{charlier,Tayo} :
\begin{eqnarray}\label{dosimpurity}
\rho(E) ={2 \over N_e \Omega} \sum_{n=1}^{N} \sum_{\lambda=v,c}\int_{-\pi/T}^{\pi/T} dk~ \delta(E - E_{\lambda n}(k) )
\end{eqnarray}
where $N_e$ is the total number of $\pi$ electrons in the GNR, $\Omega = 2\pi/N_cT$ is the length of the 1D reciprocal space for each allowed state, $N_c$ is the number of unit cells in the AGNR of finite length, $L = N_c T$, $T$ being the unit cell length. For computational purposes, we replace the Dirac delta function with a Lorentzian with the line width $\Gamma = 0.01$ eV. We present the DOS for an energy range of $\pm 2$ eV around the Fermi energy $E_F =0$.
The DOS for $N = 12, 13,$ and 14 H-passivated AGNR for three values of strain: -16\%, 0\%, and +16\% is shown in Fig. \ref{densityofstates}. For $N=12$, the band gaps for $\sigma= -16\%$ and $0\%$ are approximately equal, while the band gap shrinks for $\sigma = +16\%$. The peak of the first van Hove singularities (VHSs) is approximately the same for all three values of strain. However, the peaks tend to build up for tensile strain, compared to compressive strain. For $N=13$, the band gap decreases as the strain changes from compressive to tensile. For $N =14$, the band gap shrinks as $\sigma$ changes from -16\% to 0\%, then expands to approximately its original value when $\sigma = +16\%$. Generally, the positions of the VHSs change with applied strain and the peak heights get enhanced for positive strain, for the range of energy considered.

\section{Conclusion}
In summary, we have shown that edge passivation and the presence of strain can both be described by the same model Hamiltonian within the TB model simply by renormalizing the C-C hopping integral. We calculated the electron mass versus band gap energy for strained H-passivated AGNRs belonging to three families: $N = 3p, 3p+1 , 3p+2$. For unstrained H-passivated AGNRs, the effective mass exhibits a linear dependence on band gap energy for small energy gaps or large ribbon width. However for ribbons with small width or larger band gaps, the effective mass dependence on energy gap is parabolic. Analytic fits were also obtained for AGNRs belonging to different families. The effect of H-passivation on the effect mass is very prominent for ribbons with small widths. In the presence of strain, both band gap and effective mass displays a nearly zigzag periodic pattern, indicating that the effective mass remains proportionate to the band gap even in the presence of applied strain. Our analysis provides further insights into the uniqueness of graphene's electronic properties and is useful for studying carrier mobility in intrinsic AGNRs semiconductors where carrier scattering by phonons is the dominant scattering mechanism.

\section*{Acknowledgement}
The author acknowledges support from the department of physics, Pittsburg State University.

\end{document}